\documentstyle[12pt]{article}  
\begin{document}  
  
\def\bce{\begin{center}}  
\def\ece{\end{center}}  
\def\beq{\begin{eqnarray}}  
\def\eeq{\end{eqnarray}}  
\def\ben{\begin{enumerate}}  
\def\een{\end{enumerate}}  
\def\ul{\underline}  
\def\ni{\noindent}  
\def\nn{\nonumber}  
\def\bs{\bigskip}  
\def\ms{\medskip}  
\def\tr{\mbox{tr}}  
\def\wt{\widetilde}  
\def\wh{\widehat}  
\def\brr{\begin{array}}  
\def\err{\end{array}}  
\def\dsp{\displaystyle}  
\def\eg{{\it e.g.}}  
\def\ie{{\it i.e.}}

\vspace*{-10mm}  
  
\hfill IEEC/CSM-99-61  
  
\hfill hep-th/9906229  
  
\hfill June 1999 
 
\thispagestyle{empty}

\vspace*{4mm}

\begin{center}  
  
{\LARGE \bf  On the concept of determinant for the differential operators  
 of Quantum Physics}

\vspace{4mm}  
  
\medskip

{\sc E. Elizalde}\footnote{E-mail:  
elizalde@ieec.fcr.es \ eli@ecm.ub.es \ \  
http://www.ieec.fcr.es/cosmo-www/eli.html} \\  
Consejo Superior de Investigaciones Cient\'{\i}ficas (CSIC),\\  
Institut d'Estudis Espacials de Catalunya (IEEC), \\  
Edifici Nexus 201, Gran Capit\`a 2-4, 08034 Barcelona, Spain\\ and \\  
Departament ECM i IFAE, Facultat de F\'{\i}sica, \\  
Universitat de Barcelona, Diagonal 647,  
08028 Barcelona, Spain \\

\vspace{6mm}  
  
{\bf Abstract}  
  
\end{center}  
  
The concept of determinant for a linear operator in an   
infinite-dimensional space is addressed, by using the   
derivative of the operator's zeta-function (following Ray and Singer)   
and, eventually, through its zeta-function  
trace.  A little play with operators as simple as $\pm I$  
($I$ being the identity operator) and variations thereof, shows that the   
presence of a non-commutative anomaly (\ie , the fact that det $(AB) \neq$ det   
$A$ det $B$), is unavoidable, even for commuting  and, remarkably, also for  
 almost constant  
operators. In the case of Dirac-type operators, similarly basic   
arguments lead to the conclusion ---contradicting common lore--- that in spite   
of being   
$\det (\slash\hspace{-3mm}D +im ) = \det (\slash\hspace{-3mm}D -im )$  
(as follows from the symmetry condition of the $\slash\hspace{-3mm}D$-spectrum),  
it turns out that these determinants may {\it not} be equal to  
$\sqrt{\det (\slash\hspace{-3mm}D^2 +m^2 )}$, simply because   
$\det [(\slash\hspace{-3mm}D +im ) (\slash\hspace{-3mm}D -im )] \neq   
\det (\slash\hspace{-3mm}D +im ) \, \det (\slash\hspace{-3mm}D -im )$.  
A proof of this fact is given, by way of a very simple example,  
 using operators with an  
harmonic-oscillator spectrum and fulfilling the symmetry condition.  
This anomaly can be physically relevant if, in  
addition to a mass term (or instead of it), a chemical potential  
contribution is added to the Dirac operator.

\vfill  
  
  
  
  
\newpage  
  
\noindent{\large \bf 1. Introduction}   
\ms  
  
Many fundamental calculations of Quantum Field Theory reduce, in essence,   
to the   
computation of the determinant of some operator. One could even venture to say   
that, at one-loop order, any such theory reduces to a theory of determinants.  
 The  
operators involved are `differential' ones, as the normal physicist   
would say. In fact, properly speaking, they are pseudodifferential operators   
($\Psi$DO), that is, in loose terms `some analytic functions of differential   
operators' (such as $\sqrt{1+D}$ or $\log (1+D)$, but {\it not!} $\log D$).  
This is explained in detail in Refs. \cite{elicmp}--\cite{psdo}.  
  
Important as the concept of determinant of a differential or $\Psi$DO may be  
for theoretical physicists (in view of what has just been said), it is   
surprising that this seems not to be a subject of study among  
function analysts or mathematicians in general. This statement must be   
qualified:  
I am specifically refering to determinants that involve in its definition  
some kind of regularization, very much related to operators that are not  
traceclass. This piece of calculus ---always involving regularization---   
falls outside the scope of the standard disciplines and even many physically oriented   
mathematicians know little or nothing about it. In a sense, the subject  
has many things in common with that of {\it divergent series} but has not been  
so groundly investigated and lacks any reference comparable to the   
very beautiful book of Hardy \cite{hardy}. Actually, from this general   
viewpoint, the question of regularizing infinite determinants was already   
addressed by Weierstrass in a way that, although   
it has been pursued by some theoretical physicists with success, is not   
without problems ---as a general method--- since it ordinarily leads   
to non-local contributions that cannot be given a physical meaning in QFT.   
We should mention, for completion, that   
there are, since long ago, well stablished theories of determinants for  
degenerate operators, for traceclass operators in the Hilbert space,  
 Fredholm operators, etc. \cite{kato}  
but, again, these definitions of determinant do not fulfill all the needs   
mentioned above which arise in QFT.  
  
Any high school student knows what a determinant is, in simple words, or at least  
 how to calculate the determinant of a $3\times 3$ matrix  
(and some of them, even that of a $4\times 4$ one). But many one prominent   
mathematician will  
answer the question: {\it What is your favourite definition of determinant   
of a differential operator?} with: {\it I don't have any}, or: {\it These   
operators don't have determinants!} An even more `simple' question I   
dare to ask the reader  
(which she/he may choose to ask to some other colleague on its turn) is the   
following:  
{\it What is the value of the determinant of minus the identity operator in an infinite  
dimensional space?} Followed by: {\it And that of the determinant $\prod_{n \in N}  
(-1)^n$?} {\it Is it actually equal to the product of the separate determinants of  
the plus 1s and of the minus 1s?}   
  
In this contribution I will point out to specific situations,  
some of them having become common lore already and other that have appeared  
recently in the literature, concerning the concept of determinant in QFT,   
and I will try to give `reasonable' answers to questions such as the last ones.  
\bs  
  
  
\noindent{\large \bf 2. Infinite series and (almost) trivial determinants}   
\ms  
  
The mathematical theory of divergent series has been very fruitful in  
taming the infinites that have appeared in QFT, from the very begining of   
its conception. Its role is very essential, at least in the first stage of the   
regularization/renormalization procedure. Euler and Borel summation methods,   
and analytic continuation techniques are there commonly used. But  
some difficulties exist that are   
inherent to the theory of divergent series (see, for instance,  
\cite{hardy}). One of them is the well known fact that, sometimes, by using   
different schemes, different results are obtained. In a well posed physical   
situation, the `right' one can then only be choosen after experimental validation.  
Another problem is to understand, in physical terms, what you are doing, while  
performing say an analytic continuation from one region of the complex  
plane to another \cite{eli2}. This has prevented \eg the zeta function regularization   
procedure from getting general aceptance among common physicists.  
  
The situation  concerning infinite determinants is even worse, in a sense.   
There is no book on the subject to be compared, for instance,  
 with the above mentioned  
one by Hardy and we see every day that dubious manipulations are being performed   
at the level of the eigenvalues, that are then translated to the determinant  
itself and elevated sometimes to the cathegory of standard results ---when  
not of lore theorems. The first problem is the definition of the determinant  
itself.  
Let me quote in this respect from a recent paper by E. Witten \cite{Witten1}:   
{\it The   
determinant of the Dirac operator is defined roughly as  
\beq  
\det {\cal D} = \prod_i \lambda_i,  
\eeq  
where the infinite product is regularized with (for example) zeta   
function or Pauli-Villars regularization}. The zeta function definition of the   
determinant   
\beq  
{\det}_\zeta {\cal D} =\exp \left[-{\zeta_{\cal D}}' (0)\right],  
\eeq  
is maybe the one that has more firm mathematical grounds \cite{RS}. In spite of  
starting from the identity: log det $=$ tr log,  it is known to develop the  
so called {\it multiplicative anomaly}: the determinant of the product of two  
operators is not equal, in general, to the product of the determinants 
(even if the operators commute!).  
This happens already with very simple operators (as two one-dimensional   
harmonic oscillators only differing in a constant term, Laplacians plus 
different mass terms, etc.). It  
may look incredible, at first sight, from the tr log property  
and the additivity of the trace, but   
we must just take into account that the zeta trace is {\it no} ordinary trace  
(for it involves regularization), namely:  
\beq  
\tr_\zeta {\cal D} =\zeta_{\cal D} (-1),  
\eeq  
so that $\tr_\zeta (A+B) \neq \tr_\zeta A + \tr_\zeta B$, in general.  
Not to understand this has originated a considerable amount of errors in  
the specialized literature ---falsely attributed to missfunctions of the   
rigorous and elegant zeta function method!  
  
As an example, consider the following commuting linear operators  
in an infinite-dimensional space, given in diagonal form by:  
\beq  
O_1 = \mbox{diag\ } (1,2,3,4, \ldots ), \qquad  
O_2 = \mbox{diag\ } (1,1,1,1, \ldots ) \equiv  I,  
\eeq  
and their sum  
\beq  
O_1 +O_2 = \mbox{diag\ } (2,3,4,5, \ldots ).  
\eeq  
The corresponding $\zeta$-traces are easily obtained:  
\beq  
&&\tr_\zeta O_1 =  \zeta_R (-1) = - \frac{1}{12},  \qquad  
\tr_\zeta O_2 =  \zeta_R (0) = - \frac{1}{2},  \nn \\ &&  
\tr_\zeta (O_1 +O_2)=  \zeta_R (-1) -1= - \frac{13}{12},  
\eeq  
the last trace having been calculated according to the rules of  
infinite series summation (see \eg , Hardy \cite{hardy}). We observe that  
\beq  
\tr_\zeta (O_1 +O_2) - \tr_\zeta O_1 - \tr_\zeta O_2=  - \frac{1}{2} \neq 0.  
\eeq  
If this happens in such  simple situation, involving the identity  
operator, one can easily imagine that any precaution one can take in  
manipulating infinite sums might  
turn out to be insufficient. Moreover, since the multiplicative anomaly  
---as has been pointed out before--- originates precisely in the  
failure of this addition property for the regularized trace, we can already  
guess that it also can show up in very simple situations, as will now be  
proven,  in fact.  
The appearance of the multiplicative anomaly prevents, in particular,  
naive manipulations with the eigenvalues in the determinant, as reorderings and  
splittings, what a number of physicists seem not to be aware of.  
  
For warming up, let us calculate some simple determinants with the zeta  
function method. To start with, take  
\beq  
\Delta_1 = \prod_{n=1}^\infty n.  
\eeq  
We have  
\beq  
\zeta_1(s)= \sum_{n=1}^\infty n^{-s} =\zeta_R (s), \qquad  {\zeta_R}' (0)  
=-\frac{1}{2} \log (2\pi),  
\eeq  
so that  
\beq  
\Delta_1 = \exp \left[\frac{1}{2} \log (2\pi)\right] =\sqrt{2\pi},  
\eeq  
a nice result. In the same way, we obtain  
\beq  
\Delta_2 = \prod_{n=1}^\infty n^{-1}= \frac{1}{\sqrt{2\pi}}  
\eeq  
---from $\zeta_2(s) =\zeta_R (-s)$--- as should be expected.  
  
Let us now consider the apparently more simple case:  
\beq  
\Delta_3 = \prod_{n=1}^\infty \lambda.  
\eeq  
This poses a problem to the zeta function method, which must be modified  
somehow to cope with such situation. In fact, the corresponding  
zeta function,  
\beq  
\zeta_3(s)= \sum_{n=1}^\infty \lambda^{-s},  
\eeq  
has no abscissa of convergence in the complex plane (since the sequence of  
eigenvalues is neither increasing nor decreasing. This can be, however,  
naturally solved as follows: by taking logarithms (what is inherent with the  
definition of the zeta function method) and using again 
the rules for infinite series, it is plain that the result is  
\beq  
\log \Delta_3 = \log \lambda \ \sum_{n=1}^\infty 1 = -\frac{1}{2} \,  
\log \lambda,   
\eeq  
where the factor in  front of $\log \lambda$ may be interpreted as the   
`zeta measure' of the set of positive natural numbers, thus  
\beq  
\Delta_3 = \prod_{n=1}^\infty \lambda = \lambda^{-\frac{1}{2}}.\label{eq11}  
\eeq  
This leads, in particular, to the following results:  
\beq  
\Delta_4 = \prod_{n=1}^\infty 1 = 1^{-\frac{1}{2}} =\pm 1,  
\eeq  
for the determinant of the identity operator, $I$, and  
\beq  
\Delta_5 = \prod_{n=1}^\infty (-1) = (-1)^{-\frac{1}{2}}=\mp i,  
\eeq  
for the determinant of the operator $-I$. As it seems clear that the determinant  
of the identity operator should be 1, this tells us (by choice everywhere   
of the {\it  
same} determination of the logarithm in the complex plane) that the determinant  
of $-I$ is $-i$ and, that of $\lambda I$, the inverse of the corresponding  
square root of $\lambda$ in (\ref{eq11}). Notice that, in this way,  
 we are starting to build up a set  
of consistency rules that are reminiscent, in some manner, of the corresponding rules  
for infinite series \cite{hardy}. More than this, by use of the logarithm,  
{\it all} the ordinary rules for infinite series are appliable to the series of  
logs of eigenvalues, in particular, the ones concerning multiplication by a  
{\it common factor} (used before already), or of splitting out a {\it finite}   
number of first  
terms from the series (that is to say, a finite number of first factors from  
the determinant). However, the splitting of an infinite number of terms ---or of  
the whole series into two--- is {\it not} allowed in general.  That is, again, the  
lesson we have learned from the existence of the multiplicative anomaly of the  
determinant when evaluated by the zeta function procedure.  
An additional comment is in order:  in  dealing with  
infinite series we always  take logarithms, and  this introduces an ambiguity  
in the zeta function definition of the determinant. This fact is well known  
\cite{Kons,Semin2} and is common to other regularization methods   
(as Pauli-Villars'), under different disguises.  
It can duely be taken care of by sticking to one and the same determination of the  
logarithm during the whole calculation.  
  
An apparent problem ---or virtue perhaps?--- of the zeta function  
definition of infinite  
determinants is posed by the following example.  
It turns out that the determinants  
\beq  
\Delta_6 = \prod_{n=1}^\infty (-1)^{2n+1}, \qquad  
\Delta_7 = \prod_{n=1}^\infty (-1)^{4n+1},\qquad \ldots  
\eeq  
and $\Delta_5$ are all different. This originates in   
\beq  
\zeta_6 (-1)&=& 2 \zeta_H(-1,1/2)-1= - B_2 (1/2)-1 =- \frac{11}{12}, \nn \\  
\zeta_7 (-1)&=& 4 \zeta_H(-1,1/4)-1= -2 B_2 (1/4)-1 =- \frac{23}{24},\quad \ldots  
\eeq  
and can be interpreted as due to the change of the 'zeta  
measure' of the number of factors in the product leading to the determinant.  
We could try to avoid this problem by sticking always to the most simple  
characterization of the eigenvalues series (in this case $-1, -1, -1,   
\ldots$, any reference  
to $n$ being superfluous). Things are, regretfully, not that simple.  
Consider the determinants $\Delta_4$, $\Delta_5$, and  
\beq  
\Delta_8 = \prod_{n=1}^\infty (-1)^n,      
\eeq  
and try to make compatible the apparently obvious fact that:   
\beq  
\Delta_8 = \sqrt{\Delta_4 \Delta_5}.   \label{sp1}  
\eeq  
The determinant  $\Delta_8$ can be obtained in three different ways, that  
yield the same result. \\  
(i) We have, through the corresponding zeta function,  
\beq  
&&\zeta_8 (s) =\sum_{n=1}^\infty (-1)^{-ns} = \frac{1}{(-1)^s -1}, \nn \\  
&&{\zeta_8}' (s) = \log (-1) \left[ \frac{1}{\pi^2s^2} + \frac{1}{12} +  
{\cal O} (s) \right],   \nn \\  
&&\Delta_8 =\exp[-\left. {\zeta_8}' (0)\right|_{reg}] =  (-1)^{-1/12}.  
\eeq  
(ii) On the other hand, taking logs as before, from the zeta function measure of the  
set of exponents, we get  
\beq  
&&\log \Delta_8 =\log (-1)\ \sum_{n=1}^\infty n =\log (-1)\zeta_R (-1) , \nn \\  
&&\Delta_8 =(-1)^{\zeta_R (-1)} =  (-1)^{-1/12}.  
\eeq  
(iii) Also, we may instead choose to take derivatives, term by term, in the first  
 expression for the zeta function  
\beq  
&&\zeta_8 (s) =\sum_{n=1}^\infty (-1)^{-ns}, \nn \\  
&&{\zeta_8}' (s) =-\sum_{n=1}^\infty n (-1)^{-ns} \log (-1), \ \   
{\zeta_8}' (0) = - \log (-1)\sum_{n=1}^\infty n =\frac{\log (-1)}{12},  
  \nn \\  
&&\Delta_8 =\exp[-{\zeta_8}' (0)] =  (-1)^{-1/12}.  
\eeq  
Remarkably enough, in all three cases we obtain the same result for this  
 determinant. Let us now  
try to fulfill the factorization condition (\ref{sp1}). We have:  
\beq  
&&\Delta_4=\prod_{n=1}^\infty  1=1^{-1/2}=\pm 1, \qquad   
\Delta_5=\prod_{n=1}^\infty  (-1) = (-1)^{-1/2}=  
\mp i, \nn \\ &&  
\Delta_8 = \prod_{n=1}^\infty (-1)^n= (-1)^{-1/12}=  
\sqrt{\Delta_4 \Delta_5}=1^{-1/4}\, (-1)^{-1/4}. \label{eq21}  
\eeq  
The only way to fulfill this property (\ref{sp1}) is to accept that:  
\beq  
\prod_{n=1}^\infty  1=-1 \qquad  !!  
\eeq  
On the contrary, if we insists (as almost everybody would agree on) that the  
determinant of the identity is 1, then we must give up the property  
that the determinant of the alternating series of eigenvalues $1, -1, 1,  
-1, \ldots$ is equal to the subdeterminant product of the 1s, times the   
subdeterminant product of the $-1$s. This is the most simple reflection  
one could ever have expected to obtain of the {\it multiplicative anomaly}    
of the determinant!   
  
If we choose to preserve, at any price,   
 the multiplication property of the determinant  
and give sense to the strange fact that det $I =-1$, then we do attain  
compatibility in Eqs. (\ref{eq21}) by setting:  
\beq  
 \prod_{n=1}^\infty  1 = e^{i\pi}, \qquad   \prod_{n=1}^\infty  
   (-1) = e^{i\pi/2}, \qquad  
\prod_{n=1}^\infty  (-1)^n = e^{3i\pi/4},    
 \qquad   \prod_{n=1}^\infty (-1)^{4n} = e^{3i\pi},\ \ldots     \label{cho1}  
\eeq  
All these are compatible zeta function definitions of the determinant  
(they can be fixed from acceptable roots of 1 or $-1$ as given by the zeta   
function exponents)   
satisfying the multiplication rule. However, it is easy to see that this  
process cannot go for ever (and thus eliminate the anomaly): the following   
dets cannot possibly fulfill the multiplicative property:   
\beq  
\prod_{n=1}^\infty (-1)^{2n+1} = (-1)^{-11/12}\neq \prod_{n=1}^\infty  
 (-1)^{2n} \,  \prod_{n=1}^\infty (-1).  
\eeq  
But maybe this is asking too much, as has been observed before, in which case  
we are still left with the compatible (albeit really weird) choice  
(\ref{cho1}).  
 
Anyhow, it is easy to check that we do not find problems in factorizations  
like  
\beq  
\prod_{n=1}^\infty (-n) = \prod_{n=1}^\infty (-1) \ \prod_{n=1}^\infty n,  
\eeq  
since  
\beq  
&&\zeta (s) =(-1)^{-s} \zeta_R (s), \qquad \zeta' (0) =-\frac{1}{2} \log (-1)-  
\frac{1}{2} \log (2\pi), \nn \\  
&& \prod_{n=1}^\infty (-n) = e^{-\zeta'(0)} = (-1)^{-1/2} \, \sqrt{2\pi}  
= \prod_{n=1}^\infty (-1) \ \prod_{n=1}^\infty n,  
\eeq  
or in the more involved one  
\beq  
\prod_{n=1}^\infty (-1)^n n = \prod_{n=1}^\infty (-1)^n \ \prod_{n=1}^\infty n,  
\eeq  
because  
\beq  
&&\zeta (s) = \sum_{n=1}^\infty (-1)^{-ns} n^{-s} =\Phi ((-1)^{-s},s), \qquad  
\zeta' (0) =\frac{i\pi}{12}+ \Phi'(0,1)= \frac{i\pi}{12}-\frac{1}{2} \log  
(2\pi), \nn \\  
&& \prod_{n=1}^\infty (-1)^n n = e^{-\zeta'(0)} = (-1)^{-1/12} \, \sqrt{2\pi}  
= \prod_{n=1}^\infty (-1)^n \ \prod_{n=1}^\infty n,  
\eeq  
being $\Phi$ the polylogarithm function. The factorization  
of the determinants holds here again, in the zeta function prescription, and  
 this  fact does not seem to be that immediate, in view of  
the last calculation as compared with what we had before.  
  
All the above considerations may sound rather trivial, but actually they  
are not,  and should be carefully  
taken into account before proceding with the sort of manipulations  
of the eigenvalues and splittings of determinants that pervade the   
specialized literature.   
\bs  
  
  
\noindent{\large \bf 3. The multiplicative anomaly for Dirac type operators}   
\ms  
  
Consider the ordinary Dirac equation for a massive spinor   
\beq  
(\slash\hspace{-3mm}D +im ) \psi =0.  
\eeq  
Usually, the determinant of the Dirac operator acting in this equation is  
obtained by using the following argument (see, \eg, \cite{ms}):  
\beq  
\det (\slash\hspace{-3mm}D +im )=\det (\slash\hspace{-3mm}D -im )=  
\left[  \det(\slash\hspace{-3mm}D^2 +m^2)\right]^{1/2}. \label{2.1}  
\eeq  
This comes about from the fact that the spectrum of the Dirac massless  
 operator  
$\slash\hspace{-3mm}D$ has the following property: {\it if $\lambda$ belongs to the  
spectrum, then so does $-\lambda$}, that is immediately obtained by use of the  
$\gamma_5$ operator. Then, it turns out that the first det in Eq. (\ref{2.1})  
 is a product of pairs of the form:  
\beq  
(\lambda +im)(-\lambda +im)= (-\lambda -im)(\lambda -im)= (\lambda  
 -im)(-\lambda -im),  
\eeq  
the last being the pairs appearing in the second det of Eq. (\ref{2.1}).  
This is an algebraic argument, but there is also the corresponding geometric 
one, trivial after representing the spectral points in the complex plane.  
Thus, the   
first equality in (\ref{2.1}) is proven, and the second seems obvious.  
  
However, due to the existence of the  
multiplicative anomaly for infinite determinants  
[namely, the fact that, in general, $\det (AB) \neq (\det A)\,(\det B)$],   
all these formulas, obtained  
by 'simple' manipulation of the eigenvalues, must be set under suspicion and  
are in need of a rigorous check. Concerning Eq. (\ref{2.1}, the second   
equality cannot be taken for granted, since it may turn out that  
\beq  
\det[(\slash\hspace{-3mm}D +im )(\slash\hspace{-3mm}D -im )] \neq  
\det (\slash\hspace{-3mm}D +im )\  \det (\slash\hspace{-3mm}D -im ).  
\eeq  
We will show below that this is indeed the case, in a very simple, parallel  
 example   
---completely under control--- that uses as operator some square root of the   
harmonic oscillator.

Indeed, consider the square root of the harmonic oscillator obtained by  
Delbourgo in Ref. \cite{delb}.  
This example  has potentially some interesting physical  
applications, for it is well known that a fermion in an external  
constant electromagnetic field has a similar spectrum (Landau spectrum).  
Exactly in the same way as when  
going from the Klein-Gordon to the Dirac equation and paying the same price of  
doubling the number of components  
(\eg, introducing spin), Delbourgo has constructed a model for which  
there exists a square root of its Hamiltonian, which is very close to the  
one for the harmonic oscillator. It is in fact different from the  
Dirac oscillator  
introduced by several other authors, corresponding to the minimal  
substitution $\vec{p} \rightarrow \vec{p} -i\alpha \vec{r}$. The main  
difference lies in the introduction now of the parity operator, $Q$. Whereas  
creation and destruction operators for the harmonic oscillator,  
$a^\pm = P\pm i X$, are non-hermitian,  
the combinations ${\cal D}^\pm=P\pm iQX$ are hermitian and  
\begin{eqnarray}  
H^\pm \equiv ({\cal D}^\pm)^2 =P^2 +X^2 \mp Q =2H_{\mbox{osc}} \mp Q.  
\end{eqnarray}  
Notice that the parity term commutes with $H_{\mbox{osc}}$. Doubling the  
components  ($\sigma_i$ are the Pauli matrices)  
\begin{eqnarray}  
P\rightarrow -i\sigma_1 \frac{\partial}{\partial x}, \qquad  
X\rightarrow \sigma_1 x, \qquad  
Q\rightarrow \sigma_2 ,  
\end{eqnarray}  
the operators ${\cal D}^\pm$ are represented by  
\begin{eqnarray}  
{\cal D}^\pm \rightarrow -i\sigma_1  
\frac{\partial}{\partial x}\pm\sigma_3 x.  
\end{eqnarray}  
  
In the sequel,  we will only consider the operator  
$\cal D\equiv\cal D^+$.  
It has for eigenfunctions and eigenvalues, respectively,  
\beq  
\psi^\pm_n (x)=\frac{-ie^{-x^2/2}}{\sqrt{2^{n+1}\,(n-1)!\sqrt{\pi}}}  
\left(\begin{array}{c}-i\left[H_{n-1}(x)\pm H_n(x)/\sqrt{2n}\right]\\  
\left[ H_{n-1} (x) \mp H_n(x)/\sqrt{2n}\right]\end{array}\right),  
\qquad\lambda_n=\pm\sqrt{2n},\:\:n\geq1,\nonumber  
\eeq  
\beq  
\psi_0 (x)=\frac{e^{-x^2/2}}{\sqrt{2\sqrt\pi}}  
\left(\begin{array}{c} 1 \\ i \end{array} \right),  
\qquad \lambda_0 =0,  
\eeq  
where the $H_n(x)$ are Hermite polinomials.  
  
The two operators we shall consider for the calculation of the anomaly are  
$A={\cal D}+V$ and $B={\cal D}-V$, $V$ being a real,  
constant potential with  $|V|<\sqrt{2}$,  
that goes multiplied with the identity matrix in the two (spinorial)  
dimensions (omitted here). Obviously, this $V$ is to be identified with   
the mass $m$, to make contact with the case of the Dirac equation that  
we had at the begining.  
  
Notice that ${\cal D}+V$ and ${\cal D}-V$ are hermitian, commuting operators.  
The multiplicative anomaly is defined as  
\beq  
a(A,B)= \log \det (AB) - \log \det A-\log \det B = {\zeta_A}'(0) + {\zeta_B}'(0)  
-{\zeta_{AB}}'(0).  
\eeq  
The zeta function for the operator $\cal D$ reads  
\begin{eqnarray}  
\zeta_{\cal D}(s)=\sum_i\lambda_i^{-s}  
=\sum_{n=1}^\infty[1+(-1)^{-s}]  
\left( \sqrt{2n}  
\right)^{-s}=[1+(-1)^{-s}] 2^{-s/2}\zeta_R(s/2),  
\end{eqnarray}  
$\zeta_R(s)$ being the usual Riemann zeta function, which has a simple pole  
at $s=1$. The zeta function for the operators ${\cal D}\pm V$ is (see,  
for instance, \cite{ecznp})  
\begin{eqnarray}  
\zeta_{{\cal D}\pm V}(s)=\sum_i(\lambda_i\pm V)^{-s}  
= \zeta_{\cal D}(s)+ \sum_{n=1}^\infty \frac{(\mp V)^n \Gamma (n+s)}{n! \,   
\Gamma (s)} \zeta_{\cal D}(n+s).  
\end{eqnarray}  
Finally, the zeta function for the operator $({\cal D}+V)({\cal D}-V)={\cal D}^2-V^2$  
is given by:  
\begin{eqnarray}  
\zeta_{{\cal D}^2-V^2}(s)=\zeta_{\cal D}(2s)+ \sum_{n=1}^\infty \frac{V^{2n}   
\Gamma (n+s)}{n! \, \Gamma (s)} \zeta_{\cal D}(2n+2s).  
\end{eqnarray}  
We see that the anomaly in this simple case can be obtained in terms of the   
derivatives of the Riemann zeta function. The important fact is that it   
turns out to be non-zero:  
\beq  
a({\cal D}+V,{\cal D}-V) = 2V^2.  
\eeq  
We should point out that this result can be obtained from the Wodzicki formula  
for the anomaly, even if  we are working in a non-compact  
manifold \cite{ecznp}. We thus have a very  
simple example of the presence of a non-trivial anomaly for operators of  
degree one in a space of dimension one (spinorial, however).  
 
One can argue that a mass term will be absorbed by renormalization and will  
finally yield no physical contribution. Notice, however, that the situation  
is much more general than the specific case considered here, which, however,  
 even in its simplicity already accounts    
for any kind of terms not depending on the space-time coordinates. Thus  
$V$ can represent ---aside from a mass term $im$--- a constant magnetic field,  
 a finite temperature term, or   
a chemical potential. Anomalous contributions of this kind cannot  be  
absorbed by renormalization (see, \eg,   
\cite{MToms}) and can acquire a direct physical meaning \cite{ecznp,ecz2}.

  
\newpage  
  
\noindent{\bf Acknowledgments}  
  
The author is indebted with Andreas Wipf   
and Sergio Zerbini for enlightening  
discussions and with the members of the Institutes  
of Theoretical Physics of the Universities of Jena and Trento,  
where the main part of this work was done, for  
warm hospitality.  
 This investigation has been supported by DGICYT (Spain), project  
PB96-0925, by CIRIT (Generalitat de Catalunya), by the Italian-Spanish  
program INFN--CICYT, and by the  
German-Spanish program Acciones Integradas, project HA1997-0053.  
 
\vspace{15mm} 


\begin{thebibliography}{99}  
  
\bibitem{elicmp} E. Elizalde,   
Commun. Math. Phys. {\bf 198}, 83 (1998).  
  
\bibitem{cald} A.P. Calder\'on and A. Zygmund,  
Am. J. Math. {\bf 79}, 801 (1957); Studia Math. {\bf 20}, 171  
(1961); A.P. Calder\'on and R. Vaillancourt, Proc. Nat.  
 Acad. Sci. U.S.A. {\bf 69}, 1185 (1972).  
  
\bibitem{psdo} L. H\"ormander, {\it The analysis of  
 partial differential operators}, Vols I-IV (Springer, Berlin,  
1983-85); F. Treves, {\it Introduction to pseudodifferential  
 and Fourier integral operators}, Vols. I and II  
 (Plenum, New York, 1980); M.E. Taylor, {\it Pseudodifferential  
 operators} (Princeton University Press, Princeton,  
1981); H. Lawson and M.L. Michelsohn, {\it Spin geometry}  
 (Princeton University Press, Princeton,  
1989).  
  
\bibitem{hardy}  
G.H. Hardy, {\em Divergent Series} (Oxford University Press, Oxford, 1949).    
  
\bibitem{kato}  
T. Kato, {\em Perturbation Theory for Linear Operators} (Springer,   
Berlin, 1980).    
  
\bibitem{eli2} E. Elizalde, {\it Ten physical  
applications of spectral zeta  
functions} (Springer, Berlin, 1995);  
E. Elizalde, S.D. Odintsov, A.  
Romeo, A.A. Bytsenko and S. Zerbini, {\it Zeta regularization  
techniques with applications} (World Sci., Singapore, 1994).  
  
\bibitem{Witten1}  
E. Witten, {\it Supersymmetric index of three-dimensional gauge theory},  
hep-th/9903005;  
E. Witten, Commun. Math. Phys. {\bf 121}, 351 (1989).  
  
\bibitem{RS}  
 D.B. Ray, Adv. in Math. {\bf 4}, 109 (1970);  
 D.B. Ray and I.M. Singer, Adv. in Math. {\bf 7}, 145 (1971);  
  Ann. Math. {\bf 98}, 154 (1973).  
  
\bibitem{Kons}  
M.~Kontsevich and S.~Vishik.  
{\it Functional Analysis on the Eve of the 21st Century}.  
volume~1,  173--197,  (1993).  
  
\bibitem{Semin2}  
D. Seminara, {\em Parity and large gauge invariance in thermal QED$_3$},  
hep-th/9812137; S. Deser, L. Griguolo and D. Seminara,  
Phys. Rev. {\bf D57}, 7444 (1998).  
  
\bibitem{ms}  
D.G.C. McKeon and C. Schubert, Phys. Lett. {\bf  
 B440}, 101 (1998).  
  
\bibitem{delb}  
R.~Delbourgo, {\em A square root of the harmonic oscillator}.  
 University of Tasmania Preprint  hep--th/9503056 (1995).  
  
\bibitem{ecznp}  
E. Elizalde, G. Cognola and  S. Zerbini,  
Nucl. Phys. {\bf 532}, 407 (1998).  
  
\bibitem{MToms}  
J.J. McKenzie-Smith and D.J. Toms, Phys. Rev. {\bf D58}, 105001 (1998).  
 
\bibitem{ecz2}  
 E. Elizalde, A. Filippi, L. Vanzo and  S. Zerbini, Phys. Rev. {\bf D57}, 
 7430 (1998); {\sl Is the multiplicative 
  anomaly dependent on the regularization?},  hep--th/9804071 (1998); 
{\sl Is the multiplicative  anomaly relevant?},  hep--th/9804072 (1998); 
J. S. Dowker,  {\sl On the relevance of the multiplicative anomaly},   
hep--th/9802029 (1998);  T. S. Evans,   
{\sl Regularization schemes and the multiplicative anomaly},   
hep--th/9803184 (1998). 
 
 
\end{thebibliography}
\end{document}